\documentclass[doublecol]{epl2} 
\usepackage{graphicx}
\usepackage{color}
\usepackage{amsmath}

\title{Nanoflows through disordered media: a joint Lattice Boltzmann and
Molecular Dynamics investigation}

\author{J. Russo\inst{1} \and J. Horbach\inst{2} \and F. Sciortino\inst{1} \and S. Succi\inst{3,4}}

\institute{
  \inst{1} Dipartimento di Fisica and CNR-INFM-SOFT, Universit\`a di Roma La Sapienza, Piazzale A. Moro 2, 00185 Roma, Italy \\
  \inst{2} Institut f\"ur Materialphysik im Weltraum, Deutsches Zentrum f\"ur Luft- und Raumfahrt (DLR), 51170 K\"oln, Germany\\
  \inst{3} Istituto Applicazioni Calcolo, CNR, via dei Taurini 19, 00185 Roma, Italy\\
  \inst{4} Freiburg Institute for Advanced Studies, Albert Ludwig University, Freiburg, Germany
}

\abstract{
We investigate 
nanoflows through dilute disordered media  by means of joint lattice Boltzmann (LB) and
molecular dynamics (MD) simulations --- when the size of the obstacles is comparable to the size
of the flowing particles  --- for randomly located spheres and for a correlated particle-gel.
In both cases at sufficiently
low solid fraction, $\Phi<0.01$, LB and MD provide similar values
of the permeability. However, for $\Phi > 0.01$, MD
shows that molecular size effects lead to a decrease of the permeability, as
compared to the Navier-Stokes predictions.
For gels, the simulations highlights a surplus of permeability, which 
can be accommodated within a rescaling of the effective radius of the gel
monomers.
}

\begin{document}

\maketitle
\section{Introduction}
Flow phenomena in disordered media are a subject of great theoretical and
practical interest \cite{vafai00,sahimi93}.  In case of a
fluid streaming through a random, low-density porous matrix, descriptions
based on continuum hydrodynamics have been provided\cite{brinkman47,kim_russel,ladd2,hill,hoef,warren}.
Continuum theories are expected to hold
on macroscopic scales and thus might be non-applicable to fluid
flow through microfluidic devices or through micro-gel matrices where
the typical size of the pores is at nanometer length scales and below.

However, as shown in several simulation studies (see
e.g.~Refs.~\cite{rapaport86,horbach_succi}), hydrodynamics often holds
down to the molecular scale, as far as simple steady state flows of dense
liquids are concerned. Similar conclusions have also been reached recently
for the non-trivial case of microflows over super-hydrophobic surfaces
\cite{sbragaglia06}.  However, in particular for the case of nanoflow 
in disordered media, the question of whether/to what extent continuum theory is applicable to
flows at the nanoscale in porous materials, remains open to this day. 
Indeed, previous studies on flow phenomena in disordered
media have employed only mesoscopic or macroscopic simulation
methods, such as finite element schemes \cite{cao98,andrade01}, the LB method
\cite{succi89,koponen98,cancelliere_succi,calisucci,spaid97,manwart02,capuani03,yiotis07},
and smoothed particle dynamics \cite{jiang08,vakilha08}. 
In this 
work, we address such a question using a combination of Lattice-Boltzmann
(LB) and molecular dynamics (MD) simulations.  While LB is used as an
effective Navier-Stokes equation solver, the MD simulations are performed
to solve Newton's equations of motion for a three-dimensional system
of soft spheres. As porous media, we consider non-overlapping random
arrangements of particles, as well as particle gel networks at different packing
fractions. In both cases, flows through bulk porous media and through
porous media confined between parallel plates are considered.
First, we show
that a {\it quantitative} mapping between LB and MD can be established.
Second, we study to what extent continuum theory is applicable at the
molecular scale. This is particularly interesting for gel networks, since
they introduce long-ranged structural correlations which are not
present in the random arrangement of obstacles. For this case,
we quantify deviations to predictions of the theory.  Interestingly,
the theory is renormalizable, i.e.~the observed deviations due to an
excess of permeability can be reabsorbed within a readjustment of the
effective radius of the gel monomers.

\section{Theory}
For low Reynolds number flow, Darcy
\cite{darcy1856} first established empirically a linear relation between
the average volumetric flow velocity through unit cross-sectional area,
$\langle \vec{v} \rangle$, and the pressure gradient $\vec{\nabla} p$
of the fluid across the porous medium, $ \langle \vec{v} \rangle = -
(k/\eta) \vec{\nabla} p$ (with $\eta$ the shear viscosity of the fluid and
$k$ the permeability). Note that the measurement or
calculation of $\langle \vec{v} \rangle$ implies an average over different
realizations of the porous medium.  In case of high dilution of spheres
of radius $R$ as a porous medium, Darcy's law reduces essentially to
Stokes' law, with a permeability $k_0 = 2R^2/9\phi$,
\cite{landau59,cancelliere_succi} 

Brinkman's theory is based on the stationary Navier-Stokes equations
\cite{landau59}, $\eta \nabla^2 \vec{v} -\vec \nabla p + \vec{F} = 0$ and
$\vec \nabla \cdot \vec{v} = 0$, with $\vec{v}$ the flow velocity and $\vec{F}$
an external force per unit volume acting on the fluid.  By setting 
$\vec{v} \equiv \langle \vec{v}
\rangle$ and by expressing the external force $\vec{F}$ through Darcy's law,
$\vec{F}=-\eta \vec{v}/k$, Brinkman's equation of motion \cite{brinkman47}
is obtained:
\begin{equation}
  \nabla^2 \vec{v} - \frac{1}{\eta}\vec{\nabla} p - \frac{1}{k} \vec{v} = 0 \; .
 \label{eq_brinkman}
\end{equation}
This equation describes the porous matrix as an effective medium that
exerts a friction on the fluid. The substitution of the external force
$\vec{F}$ by Darcy's law is expected to be valid only if the packing fraction
of the porous medium is sufficiently small (see below).

From Eq.~(\ref{eq_brinkman}), one can derive an explicit formula for
the flow between two parallel plates in presence of a porous medium.
Consider a gravitational force field $\rho g$ in $x$-direction and
two parallel plates at $z=-L/2$ and $z=L/2$. The porous medium between
the plates is represented by a random matrix of fixed non-overlapping
spheres. Then, the velocity profile is given by:
\begin{equation}
  v_x(z) = \frac{k g}{\nu} \left[ 
           1 - \frac{\cosh(z/\sqrt{k})}{\cosh(L/2\sqrt{k})}  
           \right]
\label{eq_pred2}
\end{equation}
with $\nu=\eta/\rho$ the kinematic velocity of the fluid.

For a dilute collection of non-overlapping spheres,
different expressions for the permeabiliy as a function of
the volume fraction, $\Phi \equiv 4 \pi R^3 n/3$, of the porous medium
have been investigated~\cite{brinkman47,kim_russel,ladd2,hill,hoef}.
A simple expression for the permeability over the entire range
of volume fractions has been proposed by van der Hoef et al.~\cite{hoef}
by fitting both LB~\cite{ladd2,hoef} and multipole expansion data~\cite{hill}
\begin{equation}
  k = k_0 \left[
10\frac{\Phi}{(1-\Phi)^3}+(1-\Phi)
\left( 1+1.5\sqrt{\Phi}\right) \; .
 \right]^{-1}
\label{eq_pred1}
\end{equation}
Note that the square-root of the permeability, $\sqrt{k}$, describes the
screening length of the flow field due to the interaction with the
porous medium.

Equations (\ref{eq_pred2}) and (\ref{eq_pred1}) are the main
results which our simulations on nanoscopic scales will be compared against.

\section{Methods and simulation details}
We use a standard Lattice Boltzmann model with single-step relaxation
term~\cite{benzi92,chen98,succi01}.  No-slip boundary conditions at solid surfaces
are implemented via standard bounce-back collision rules~\cite{ladd}.
Fully developed periodic flows are generated by pressure boundary
conditions~\cite{pressure_boundary}. The size of the LB D3Q19 lattice
is $(192\times48\times48)$ in lattice units. When present, slit walls
are placed parallel to the $xy$ plane with surfaces at locations $z=4$
and $z=43$.

\begin{figure}
\centering
\includegraphics[width=7cm]{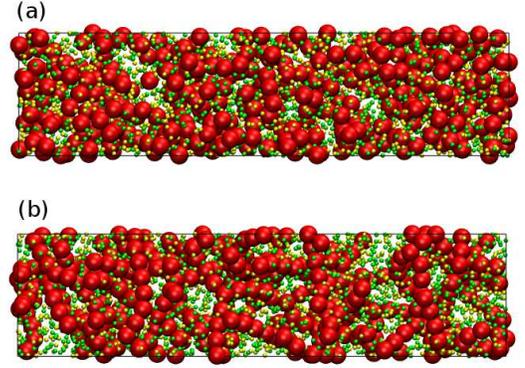}
\caption{Two snapshots at $\Phi=0.1$ with random medium (a) and
gel (b).  Porous media are coded in red, the binary mixture
fluid is coded in yellow and green.  The size of the fluid particles
is not at scale to improve visibility. \label{fig:comparison}}
\end{figure}
For the MD fluid, we choose a binary model system as proposed by
Hedges et al.~\cite{chandler} for which, also at low
temperatures, crystallization is not a problem and thus this model can
be used in forthcoming studies on glassforming fluids in porous media.

The MD fluid is composed of a 50:50 mixture of
$A$ and $B$ type particles interacting with a WCA potential,
$V_{\alpha\beta}(r)=\{4\epsilon_{\alpha\beta}[(r/\sigma_{\alpha\beta})^{-12}
-(r/\sigma_{\alpha\beta})^{-6}]+\epsilon_{\alpha\beta}\}$ for
$r<r_-=2^{1/6}\sigma_{\alpha\beta}$ (zero else). The parameters are
$\epsilon_{\rm AA}=\epsilon_{\rm AB}=\epsilon_{\rm BB}=1$ and $\sigma_{\rm AA}=1.0$,
$\sigma_{\rm AB}=11/12$, $\sigma_{\rm BB}=5/6$ \cite{chandler}.  In the following, length
and energy scales are measured in units of $\sigma=\sigma_{\rm AA}$ and
$\epsilon=\epsilon_{\rm AA}$, respectively; temperature is in units of
the potential depth $\epsilon$. The masses of both A and B particles
are taken equal, $m=m_{\text{A}}=m_{\text{B}}=1$.  The equations of motion
are integrated with the velocity form of the Verlet algorithm using a time
step $\delta t = 10^{-3}$ in units of $t_0=[m \sigma^2/\epsilon]^{1/2}$.
Thermalization at the constant temperature $T=5.0$ is obtained with a
Lowe thermostat~\cite{lowe}, which provides local momentum conservation and thus
it preserves the correct hydrodynamic behavior of the fluid.  The size of
the MD simulation box is $(32\times8\times8)$ in units of $\sigma$.
The total density of all systems is set
to $\rho=1.3$. First the porous material is introduced in the simulation
box which is then filled with the appropriate number of fluid particles
to reach the desired density.
A pressure drop is applied along the $x$ direction by adding a gravitational
field ($g$) on the fluid particles, whose intensity ($g$ in units of $\sigma/t_0^2$)
is chosen such that a linear response of the system is provided.
The kinematic viscosity $\nu$ of the fluid has been calculated aside
from separate Poiseuille-flow simulations, which yields
$\nu=24\,\sigma^2/t_0$.

To simulate slit walls (whenever present), the system is first equilibrated in a $(32\times8\times10.2)\,\sigma$ box
and then all fluid particles within a distance $z_w=1.2\sigma$ from the planes $z=0$ and $z=10.2\sigma$, are labelled as wall
particles. Wall particles retain the same interactions as fluid particles, but their position is kept fixed. In this way we account for the walls roughness and avoid layering effects.
To avoid diffusion of fluid particles in the wall region, an external potential (previously defined as $V_{AA}$)
is added along two planes parallel to the $xy$ plane, at a distance $z_p=2^{1/6}\sigma$ from the slit walls.
In this way, fluid particles feel the external potential only when they start diffusing into the walls.

To compare LB and MD, units have to be scaled appropriately. One
of our goals is to assess the validity of the LB predictions
for microscopic flows which can be resolved through MD. As shown
recently~\cite{horbach_succi}, in order to recover quantitative agreement
with MD results, the LB simulation must be taken down to microscopic
resolution, i.e.~fractions of the range of molecular interactions.
The space conversion proceeds as follows. In LB simulations the lattice
spacing, $\Delta x$ is set to unity, while in MD simulations the unit
of length is fixed by the parameter $\sigma$.  To resolve fractions of
the interaction potential we set $\Delta x<\sigma$, specifically we
choose $\sigma=6\Delta x$. This fixes the conversion of space units.
It also fixes the radius of the LB spheres as $R=\sigma/2= 3 \Delta x$.
In the rest of the letter we adopt $\sigma=\sigma_{AA}=1$ as our unit of length.

The time conversion is determined from kinematic viscosity $\nu$. The
kinematic viscosity in the LB simulation is given by $\nu_{\rm
LB}=\widetilde{\nu}_{\rm LB}\frac{\Delta x^2}{\Delta t_{\rm LB}}$,
with $\widetilde{\nu}_{\rm LB}=1/6$, while for MD we have $\nu_{\rm
MD}=\widetilde{\nu}_{\rm MD}\frac{\sigma_{\rm AA}^2}{\Delta t_{\rm
MD}}$, with $\widetilde{\nu}_{\rm MD}$ given by the Poisseuille flow
comparison.  By imposing $\nu_{\rm LB}=\nu_{\rm MD}$ and remembering
the space conversion factor we obtain for time scales $\Delta
t_{\rm MD}=\frac{\Delta t_{\rm LB}}{6^2} \frac{\widetilde{\nu}_{\rm
MD}}{\widetilde{\nu}_{\rm LB}}$.

The Reynolds number is $Re=<\vec{v}>R/\nu$, where~$<\vec{v}>$ is the spatially
averaged velocity, $R$ is the radius of the obstacles and $\nu$ is the
kinematic viscosity. We use very low external fields ($g=0.05$ in units
of $\sigma/\Delta t_{MD}^2$),
so that $Re<10^{-3}$ and simulations can be
considered under effectively zero-Reynolds number conditions.

MD and LB simulations are best compared in terms of dimensionless
quantities, such as the normalized permeability $k/k_0$, where 
$k_0$ is the permeability of a single sphere (as given by Stokes law).
Other dimensionless quantities used in the present work
are the dimensionless position $z/L$, with $L$ the width of the slit,
and the dimensionless velocity profile $u=\nu v/(gR^2)$.

Simulations of fluid flow through two types of obstacles are considered:
random media and gel media.

Random media are modeled as a collection of non-overlapping spheres of
radius $R$.  We average over $50$ independent random configuration for
each volume fraction $\Phi$ considered, in the range $\Phi=0.008\div 0.27$.

Gel networks are characteristic random structures with long range
spatial correlations.  A gel is usually made of a network of polymer
strands or from self-assembled colloidal particles.  The number of bonds
is so high that it is always possible to move from one gel-forming
monomer to another without ever leaving the network. This interlinked
structure confers the gel its peculiar properties, sharing
characteristics of both liquids and solids. It behaves like a liquid
since it can be made up primarily of fluid and allow both diffusive and
convective transport through its volume. On the other hand, a gel
is also able to support a shear stress and behave elastically, acting
like a solid.  In the present study, we neglect the elastic behaviour
of gels, keeping the monomers fixed and taking advantage of the rigid
framework through which mass transport can occur. Gel structures are
obtained through equilibrium MD simulations of Patchy particles. Patchy
particles are a class of short-ranged valence-limited particles
which can reach low temperatures without encountering the gas-liquid 
phase separation~\cite{bianchi}.
The corresponding low $T$ arrested states are in fact equilibrium gels which we use
for the present study. We follow the procedure described in~\cite{johnnew},
for networks with average valence $2.25$, equilibrated until all particles belong
to the same spanning cluster. Here, we generated 50 independent gel configurations at each
of the packing fractions $\Phi=0.03$, 0.05, 0.075, 0.1, 0.15, and 0.2.
Figure~\ref{fig:comparison} shows snapshots of MD configurations
with the random medium (a) and the gel (b) at $\Phi=0.1$.

\section{Results}
All obstacles considered in the present paper are collections of spheres
whose hydrodynamic radius is estimated by measuring the drag force
on a single sphere with periodic boundary conditions and comparing
the result with the theoretical expression given by Hashimoto\cite{hashimoto}.
For LB, this procedure leads to the an hydrodynamic radius of
$R=3.05$, slighty larger than the nominal radius (in agreement
with previous studies \cite{ladd}).
For MD we obtain a value of $R=0.44$, assuming
slip boundary conditions. The same value, $R=0.44$ can be estimated from
the interaction potential for the MD particles: at $2\cdot R=0.88$
the potential energy between two particles, separated by a distance
$2\cdot R=0.88$, is approximately equal to $k_BT$. We conclude that
for our atomistic fluid slip boundary conditions provide a consistent description
of the interactions between obstacles and fluid particles. We remind 
that the difference between
stick and slip boundary conditions is included in the Stokes expression:
$k_0=2R^2/(9\Phi)$ (stick) and $k_0=R^2/(3\Phi)$ (slip).

%
\begin{figure}
\centering
\vspace*{0.4cm}

\includegraphics[width=7cm]{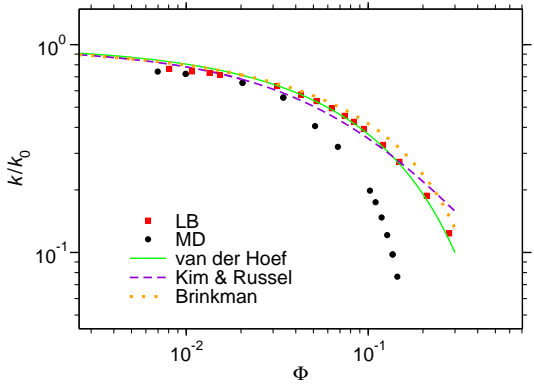}
\caption{Dimensionless permeability as a function of volume fraction
$\Phi$ for the random sphere matrix from MD simulations (circles) and
LB simulations (squares). The LB data is well described by the theoretical
prediction of Eq.~(\ref{eq_pred1}) (continuous line), except at low
volume fractions where the agreement holds with the asymptotically exact result of 
Kim and Russel~\cite{kim_russel}
(dashed line). For comparison also the result of the
Brinkman theory~\cite{brinkman47} (dotted line) is shown.
MD results consistently deviate from hydrodynamic predictions for
$\Phi>0.01$.
\label{fig:brinkman}}
\end{figure}
Figure~\ref{fig:brinkman} shows the results for the flow through random
porous media (without slit walls) for both MD and LB simulations.
LB results confirm Eq.~(\ref{eq_pred1}), except for the lowest volume
fractions,
where the low-density result of result of Kim and Russel~\cite{kim_russel}
holds. Note that the agreement with Brinkman theory is also good for low
volume fractions, say $\Phi < 0.08$.
The grid-independence of LB results
was checked by doubling the resolution ($\sigma=12\,\Delta x$).
MD data shows instead significant deviations from the hydrodynamic solution.
While for $\Phi<0.01$ the agreement between MD and LB seems to hold
(within statistical accuracy)
,
at higher volume fractions the MD results show visible
under-deviations. These can be interpreted as genuine atomistic effects,
where the reduced permeability is due to the finite size of the molecules.
Indeed, at $\Phi=0.01$, the ratio $d/\sigma$ between the average intermolecular
distance and the molecular effective diameter is of the order of $d/\sigma
\sim 4$, whereas for $\Phi=0.1$, one has $d/\sigma \sim 2$.

\begin{figure}
\centering
\includegraphics[width=7cm]{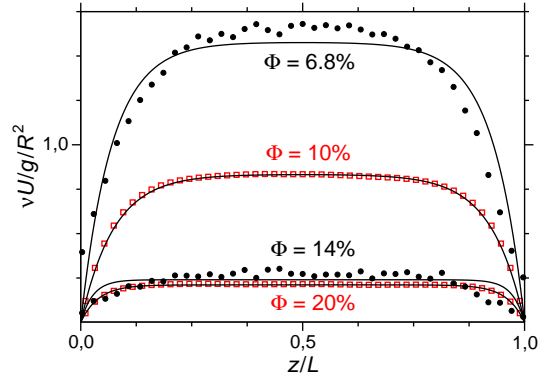}
\caption{Comparison of LB (squares) and MD (circles) for the random
medium at different volume fractions.  Full lines are the theoretical profiles,
Eq.~(\ref{eq_pred2}). Agreement between LB data and theoretical predictions
is perfect for all volume fractions investigated. MD solutions display
instead a different pattern. \label{fig:random_profiles}}
\end{figure}
We have also investigated the effect of slit walls on the permeability
$k$. To this end, we have
determined the velocity profiles
$v_x(z)$, averaged over all crossflow $yz$ sections.  Inspection of
the LB and MD velocity profiles in Fig.~(\ref{fig:random_profiles})
for random media at different values of $\Phi$ shows excellent agreement
with the prediction of Eq.~(\ref{eq_pred2}) for LB.
In contrast to that, the MD profiles show a different
pattern from the hydrodynamic solution: again, this indicates the
differences between hydrodynamic and atomistic flow for $\Phi>0.01$.

We can conclude that both global (permeability) and local (velocity
profiles) properties of an atomistic flow show consistent deviations
from the hydrodynamic solution also at relatively low volume fractions.
The results obtained show the limits of applicability of continuum approaches
when dealing with microscopic and structured fluids.

\begin{figure}
\centering
\vspace*{0.5cm}
\includegraphics[width=7cm]{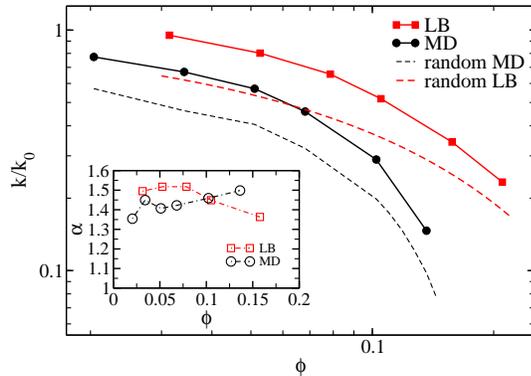}
\caption{Dimensionless permeability as a function of volume fraction
$\Phi$ for the gel matrix from MD simulations (circles) and
LB simulations (squares). The dashed line represents the results for the random
sphere matrix from Fig.~\ref{fig:brinkman}. The inset shows the
$\alpha$ parameter defined in Eq.~\ref{eq_correction} for MD (circles)
and LB (squares).}
 \label{fig:gels}
\end{figure}

We now turn to the study of the effects of the correlation between
the obstacles' positions on the flow properties for both
the hydrodynamic solution (LB) and the microscopic flow (MD).
Differently from random spheres, gel media is
characterized by short range correlations between the particles' positions.
Figure~\ref{fig:gels} reports the results for the gel media for
both MD and LB simulations.  This figure indicates that
the permeability of the gel is always higher than that of the random
media for both hydrodynamic and microscopic flows.
Also for the gel, the MD permeability
always underestimates the LB solution at packing fractions $\Phi>0.01$.

Next, we show that the increased permeability induced by the correlations
between the obstacle particles can be taken into account by
simple renormalization of the theory for random obstacles.
We first observe that the permeability can be formally written as
$k=R^2f(\Phi)$, where $f(\Phi)$ is a dimensionless function of $\Phi$.
Given the proper expression for the random sphere case [Eq.~(\ref{eq_pred1}) for
the LB data], a formal generalization for the gel case can be obtained by introducing an
effective parameter $\alpha(\Phi)$,
\begin{equation}
\label{eq_correction}
k=\alpha(\Phi)R^2f(\Phi) \; .
\end{equation}
The introduction of the parameter $\alpha(\Phi)$ formally corresponds to
the definition of an effective Stokes radius, $R_{\text{eff}}(\Phi)
= \sqrt{\alpha} R$, which depends on density. The inset of
Fig.~\ref{fig:brinkman} shows the value of the parameter $\alpha$
calculated both for LB
(squares) and MD (circles) simulations. 
For both types of simulations the effective Stokes radius remains
almost constant at all considered volume fractions, so that 
we can conclude that the gel structure results in an increase of the bare
permeability by about 50\%, which can be also interpreted as an increase
of the effective Stokes radius by a factor $1.25$.

\section{Summary and conclusions}
We have investigated 
nanoflows through disordered media by means of joint LB and MD
simulations.  For random media at $\Phi<0.01$, both LB and MD provide similar values of the permeability,
confirming that the hydrodynamic approach holds down to the nanoscale at
sufficiently low volume fractions.
For higher volume fractions, the
MD simulations reveal under-departures from the hydrodynamic solution.  Since LB
simulations still agree with the Eq.~\ref{eq_pred1} prediction, these under-departures
are most naturally interpreted as genuine atomistic effects, i.e.~breakdown
of the continuum hydrodynamic hypothesis at the nanoscale.
The same conclusions result from the study of the velocity profiles within
solid slabs filled with random media, showing that the MD flow is qualitatively
different from LB predictions.
We have then explored the effects of the correlations in the obstacles positions
by studying the flow through a correlated medium, i.e. gels.
We have shown that gels exhibit a surplus of permeability.  This surplus, about $50\%$,
can be reinterpreted as an increase of the effective radius of the gel
monomers, thereby indicating that, at least for the cases explored in
this work, the hydrodynamic solution appears to be renormalizable.

\acknowledgments
We thank A. Griesche for a critical reading 
of the manuscript. JR and FS acknowledge support from ERC 226207. JH thanks University of Roma La Sapienza for a visiting professorship.

\bibliographystyle{eplbib}

\end{document}